\documentclass[conference]{IEEEtran}
\IEEEoverridecommandlockouts
\usepackage{cite}
\usepackage{amsmath,amssymb,amsfonts}
\usepackage{algorithmic}
\usepackage{graphicx}
\usepackage{textcomp}
\usepackage{xcolor}
\usepackage{orcidlink}
\usepackage{subfig}
\usepackage{multirow}
\usepackage{xspace}
\newcommand{\X}{$\times$\xspace}    

\newcommand{\figref}[1]{\mbox{Fig.~\ref{#1}}}
\newcommand{\tblref}[1]{\mbox{Table~\ref{#1}}}

\def\BibTeX{{\rm B\kern-.05em{\sc i\kern-.025em b}\kern-.08em
    T\kern-.1667em\lower.7ex\hbox{E}\kern-.125emX}}
\begin{document}

\title{\fontsize{20}{22}\selectfont \textbf{Bhasha-Rupantarika}: Algorithm-Hardware Co-design approach for Multilingual Neural Machine Translation

\thanks{
\textsuperscript{\textdagger, \textdaggerdbl}Both authors contributed equally to this work.\\
This work was supported by the Special Manpower Development Program for Chip to Start-Up (SMDP-C2S), the Ministry of Electronics and Information Technology (MeitY), Government of India, Grant: EE-9/2/21 - R\&D-E}

 \author{
    \IEEEauthorblockN{
Mukul Lokhande\textsuperscript{\textdagger},\IEEEauthorrefmark{1}\orcidlink{0009-0001-8903-5159},
Tanushree Dewangan\textsuperscript{\textdagger},\IEEEauthorrefmark{1}\orcidlink{0009-0009-3889-1228},
Mohd Sharik Mansoori\textsuperscript{\textdaggerdbl},\IEEEauthorrefmark{1}\orcidlink{0009-0002-5274-3394},\\
Tejas Chaudhari\textsuperscript{\textdaggerdbl},\IEEEauthorrefmark{1}\orcidlink{0009-0003-3317-1375},
Akarsh J.,\IEEEauthorrefmark{1}\orcidlink{0009-0000-0376-8304},
Damayanti Lokhande\IEEEauthorrefmark{5}, 
Adam Teman\IEEEauthorrefmark{4}\orcidlink{0000-0002-8233-4711}, Senior Member, IEEE, \\
Santosh Kumar Vishvakarma\IEEEauthorrefmark{1}\orcidlink{0000-0003-4223-0077}, Senior Member, IEEE.}
\IEEEauthorblockA{\IEEEauthorrefmark{1}NSDCS Research Group, Dept. of Electrical Engineering, Indian Institute of Technology Indore, India.\\
\IEEEauthorrefmark{4}EnICS Labs, Faculty of Engineering, Bar Ilan University, Ramat Gan 5290002, Israel.\\
\IEEEauthorrefmark{5}Independent Researcher.
     }
     Email: skvishvakarma@iiti.ac.in \textbf{(Corresponding Author)}
}    

}

\maketitle

\begin{abstract}
This paper introduces Bhasha-Rupantarika, a light and efficient multilingual translation system tailored through algorithm-hardware codesign for resource-limited settings. The method investigates model deployment at sub-octet precision levels (FP8, INT8, INT4, and FP4), with experimental results indicating a 4.1\X reduction in model size (FP4) and a 4.2\X speedup in inference speed, which correlates with an increased throughput of 66 tokens/s (improvement by 4.8\X). This underscores the importance of ultra-low precision quantization for real-time deployment in IoT devices using FPGA accelerators, achieving performance on par with expectations. Our evaluation covers bidirectional translation between Indian and international languages, showcasing its adaptability in low-resource linguistic contexts. The FPGA deployment demonstrated a 1.96\X reduction in LUTs, a 1.65\X decrease in FFs, resulting in a 2.2\X enhancement in throughput compared to OPU and 4.6\X compared to HPTA. Overall, the evaluation provides a viable solution based on quantization-aware translation along with hardware efficiency suitable for deployable multilingual AI systems. The entire codes and dataset for reproducibility are  \href{https://anonymous.4open.science/r/Bhasha-Rupantarika-4678/}{publicly available}, facilitating rapid integration and further development by researchers.

\end{abstract}


\section{Introduction}
The increasing integration of artificial intelligence (AI) has heightened the emphasis on endowing smart machines with the capacity to understand, interpret, and generate human-like language, effectively merging computational capabilities with human communicative demands. Natural Language Processing (NLP) has been instrumental in this regard, encompassing a wide array of tasks such as text classification, sentiment analysis, speech recognition, among others. Multilingual Neural Machine Translation (MNMT) has attracted notable interest due to its proficiency in executing automatic, context-aware, end-to-end translations using sophisticated deep learning methodologies~\cite{TPDS, Algo-HW}. In contrast to traditional Statistical Machine Translation (SMT) systems, which depended on a dictionary-like approach through discrete components like feature extractors, translation rule extractors, and word aligners. While earlier methods utilized Convolutional Neural Networks (CNN) or Recurrent Neural Networks (RNN) under this framework, the rise of the \textbf{``Embed-Encode-Attend-Decode''} model based on the transformer framework has received considerable attention.

\begin{figure}[!t]
    \centering
    \subfloat[]{\includegraphics[width=0.95\columnwidth]{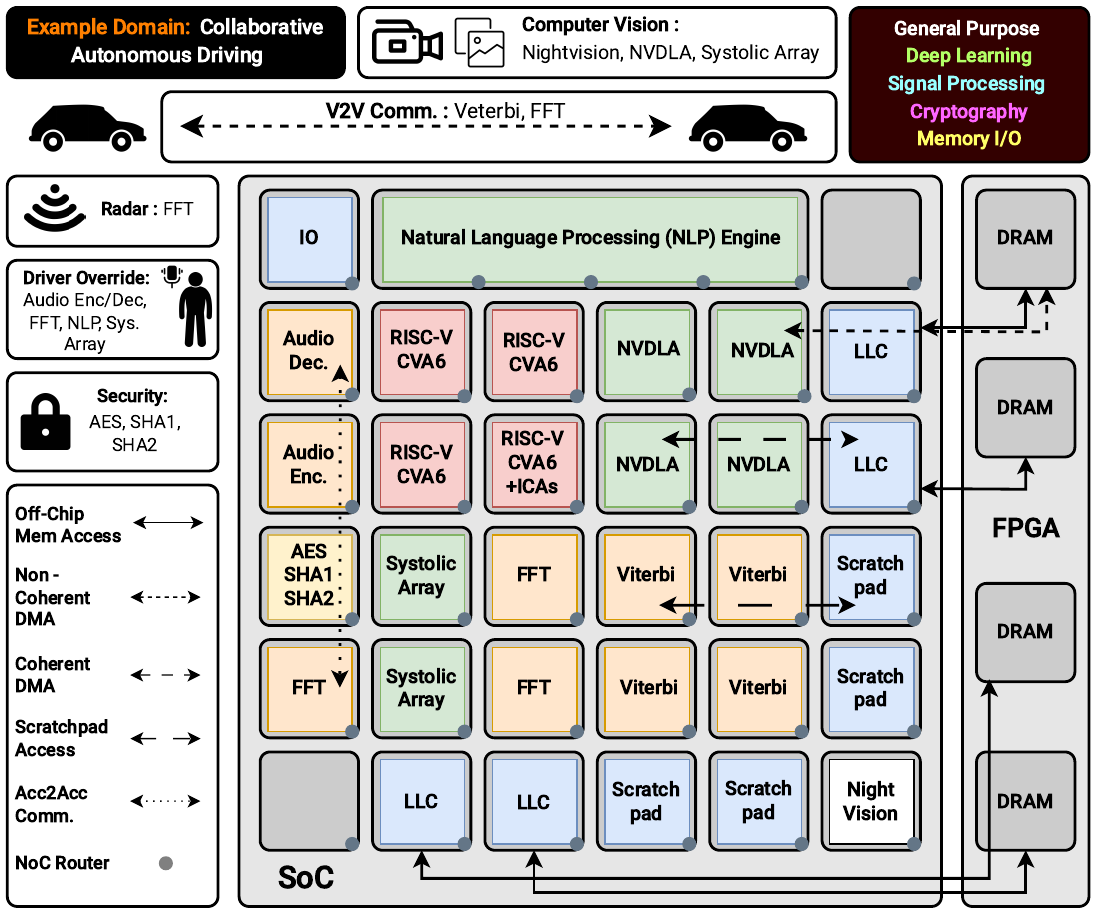}
    \label{convsoc}}
    \hfill
    \subfloat[]{\includegraphics[width=0.46\columnwidth]{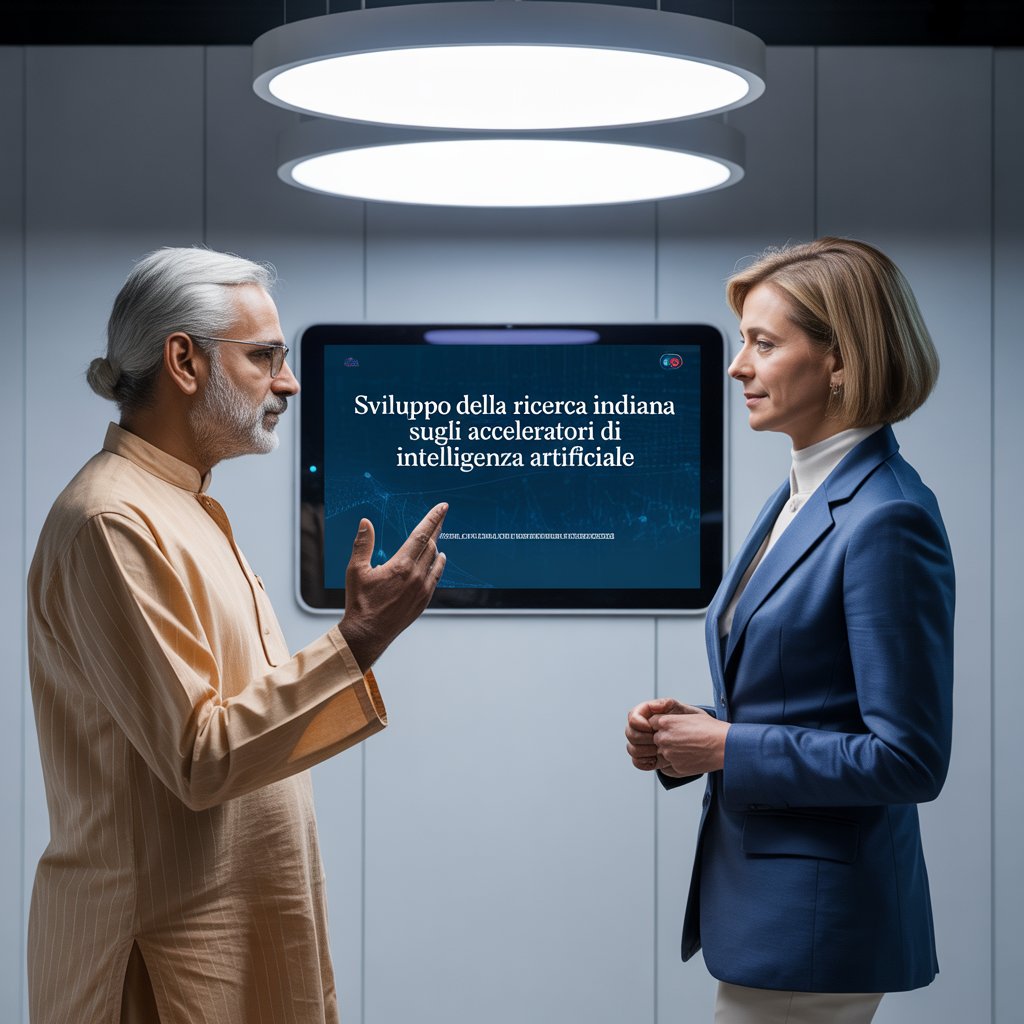}}
    \centering
    \subfloat[]{\includegraphics[width=0.46\columnwidth]{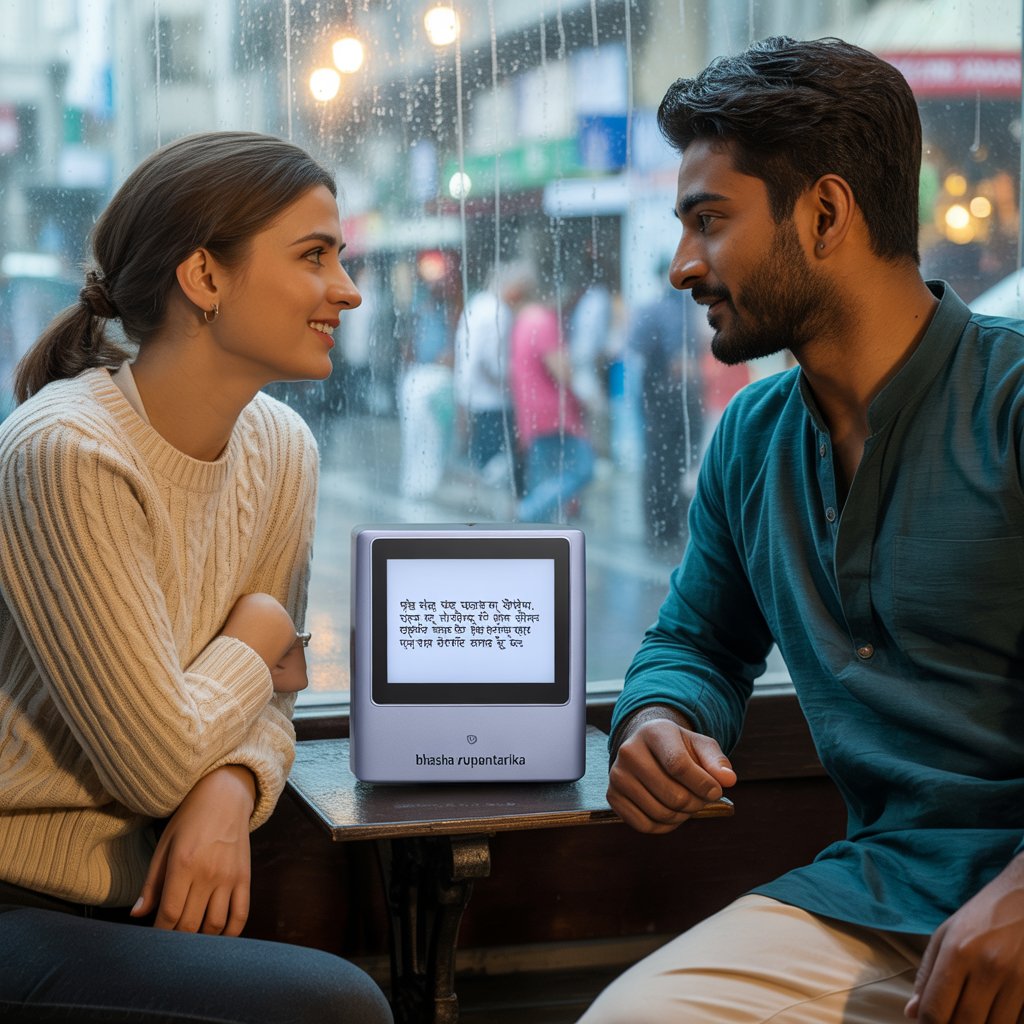}}
    
    \caption{ (a) Conventional system-on-chip (SoC) architecture, showcasing 14 domain-specific accelerators targeting multiple application domains (including NLP Engine), (b) Real-life scenario showcasing the emphasis on language translation, Indian-Italian Diplomats interaction, and (c) Two overseas students discussing.}
\end{figure}

MNMT systems are regarded as more practical because training across varied language pairs improves translation quality for less-resourced languages and aids in acquiring additional linguistic insights, a concept termed translation knowledge transfer~\cite{Raj-survey, Raj2_acm}. This approach produces a more compact and multi-faceted translation model that requires less computational power, an essential consideration for devices with limited resources. This study concentrates on translating between Indian and international languages, with the goal of implementing simple and cost-effective FPGA solutions in rural Indian areas. The importance of this work is particularly pronounced for large populations with limited formal education who nonetheless need to engage with foreign languages, especially in the fields of tourism or business.

The efficiency of resources in MNMT systems becomes particularly significant when dealing with low-resource translation scenarios. Conventionally, a single encoder is employed for multiple languages, while each target language utilizes its own separate decoder~\cite{Raj-survey}. In addition, language divergence can be mitigated by aligning commonly represented words and sentences across different languages, though this alignment requires a comprehensive grasp of multilingual representations~\cite{Raj_acm2, Survey, Indic-ST}. Transfer learning across related languages might be enhanced through the fine-grained clustering of encoder representations based on language similarity. For example, representation invariance tends to decrease on the decoder side while it increases in the upper levels of the encoder. Thus, the decoder must effectively manage representations that are specific to a language versus those that are language-neutral. Another significant factor is lexical transfer on the source side, which involves mapping pre-trained monolingual word embeddings from both parent and child languages into a unified vector space~\cite{OPU, LSTM-NLP}. 

The core question of this study is: \textbf{Is it feasible to create a single model that manages all language pairs based on the use case?} and extends to another important question: \textbf{Can shared multilingual machine translation be achieved with a low-resource NLP mechanism?} The answer is affirmative, as a unified model is indeed possible. However, it encounters challenges in representation learning and translation quality across languages, which are influenced by factors such as data precision, architecture, and learning strategies. For low-resource inference, there must be a careful equilibrium between performance and accuracy.

\begin{figure}[!t]
    \centering
    \includegraphics[width=0.825\columnwidth]{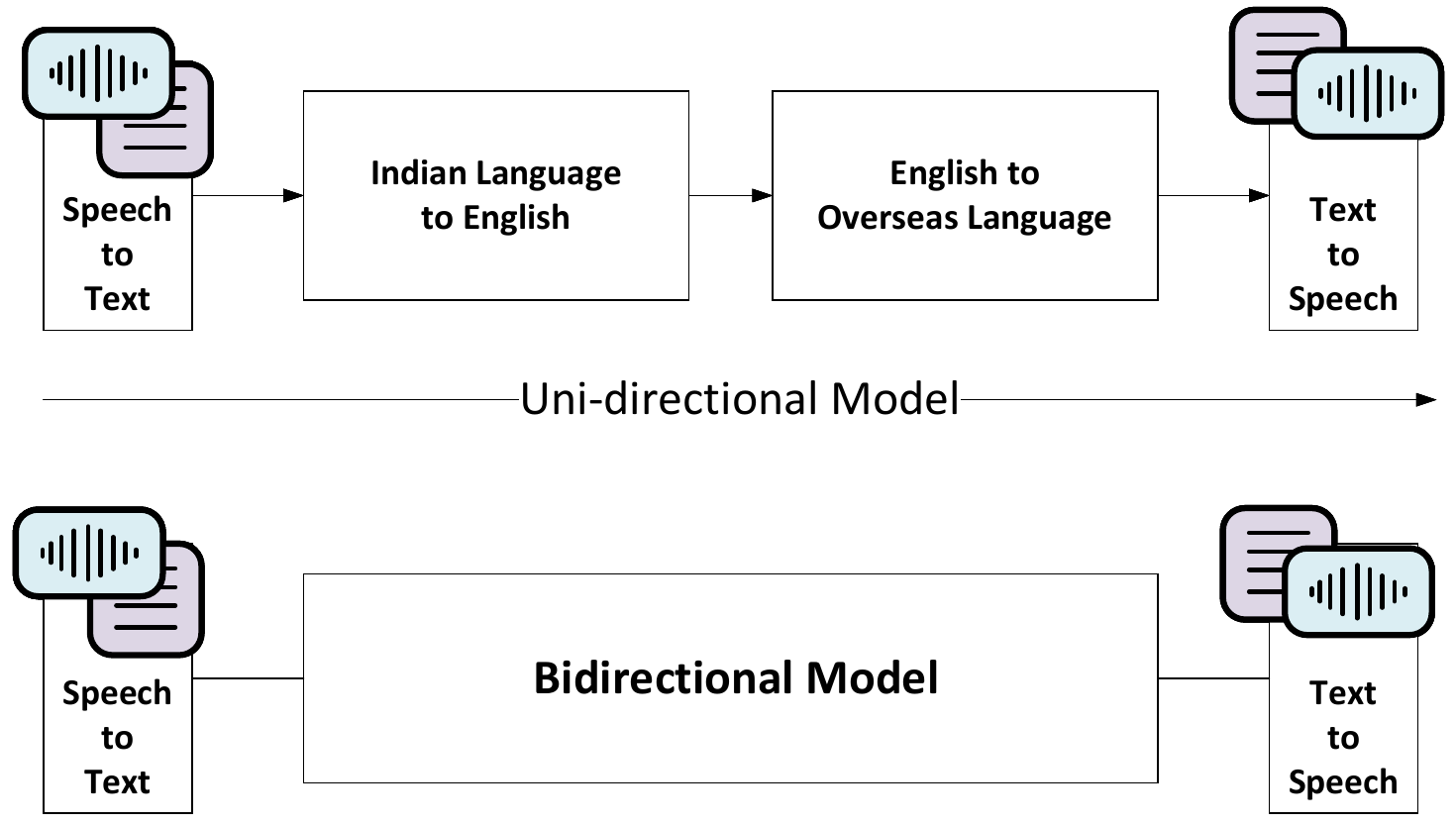}
    \caption{High level algorithm pipeline description, comparison with (a) prior approach, (b) this work for Indian language to Overseas language translation.}
    \label{fig:NLP-model}
\end{figure}

\figref{convsoc} illustrates a multi-application domain SoC designed for the concurrent execution of various computational kernels, including those from NLP, deep learning, signal processing, and cryptography. This highlights that the NLP engine is an often overlooked area in modern AI hardware developments. Consequently, the objective of this study is to tackle this issue within the Indian context. Our approach seeks to achieve translation between Indian languages and foreign languages using a unified model, differing from previous methodologies that employed separate models for translating from Indian languages to English and then from English to foreign languages, with two additional one-directional models for reverse translation. This unified approach enhances resource efficiency on edge platforms.

\section{The proposed Approach}

\subsection{Algorithm Pipeline}  

The conventional baseline employs OpenAI Whisper Large-v3 for speech-to-text conversion in foreign languages and relies on AI4Bharat IndicConformer for Indian languages. Additionally, text conversion from Indian languages to English can be achieved using IndicTrans2~\cite{bhasaanuvaad}, a distilled INT4 version of the Bhasha-anuvaad model, and translation from English to other foreign languages is done via Meta LLaMA 3.2, followed by text-to-speech (TTS) using Coqui XTTS v3 for foreign languages and IndicParler TTS~\cite{bhashini} for Indian languages. The proposed pipeline incorporates:
\begin{itemize}
    \item NLLB-200~\cite{NLLB}, a distilled version of INT4 with 600M parameters, which are lightweight transformers suitable for efficient FPGA implementation. 
    \item OpenAI Whisper Large-v3~\cite{whisper}, a recent multilingual automatic speech recognition (ASR) model, supporting over 90 languages and excelling in transcribing speech despite background noise, accents, and various recording conditions, making it ideal for real-time transcription, subtitling, and cross-lingual applications in foreign languages. 
    \item AI4Bharat IndicConformer~\cite{bhasaanuvaad}, an ASR model designed for Indian languages, providing transcription across multiple Indic languages such as Hindi, Tamil, Telugu, and Kannada, balancing speed and accuracy to accommodate India's linguistic diversity, thus serving applications in government, education, and healthcare sectors within India. 
    \item Coqui XTTS v3, a multilingual text-to-speech system with robust support across global languages, facilitating cross-lingual voice transfer by replicating one speaker's voice in another language, highly effective for dubbing and personalized speech synthesis. 
    \item IndicParler TTS, which synthesizes natural-sounding speech in Indic languages, addressing tonal, phonetic, and prosodic variations, and finds uses in voice assistants, e-learning platforms, and digital government services. 
    \item IndicTrans2 (Distilled INT4 version), a neural machine translation model with reduced size while maintaining accuracy, supporting translation between Indian languages and between Indian and English, thereby enhancing digital inclusivity in India. 
    \item Meta’s LLaMA 3.2, a large language model used for advanced reasoning and multilingual understanding, facilitating general-purpose tasks like summarization, dialogue, and content generation in foreign languages. 
\end{itemize}
    
\begin{figure}
    \vspace{-2mm}
    \centering
    \subfloat[]{\includegraphics[width=0.4\columnwidth, height=55mm]{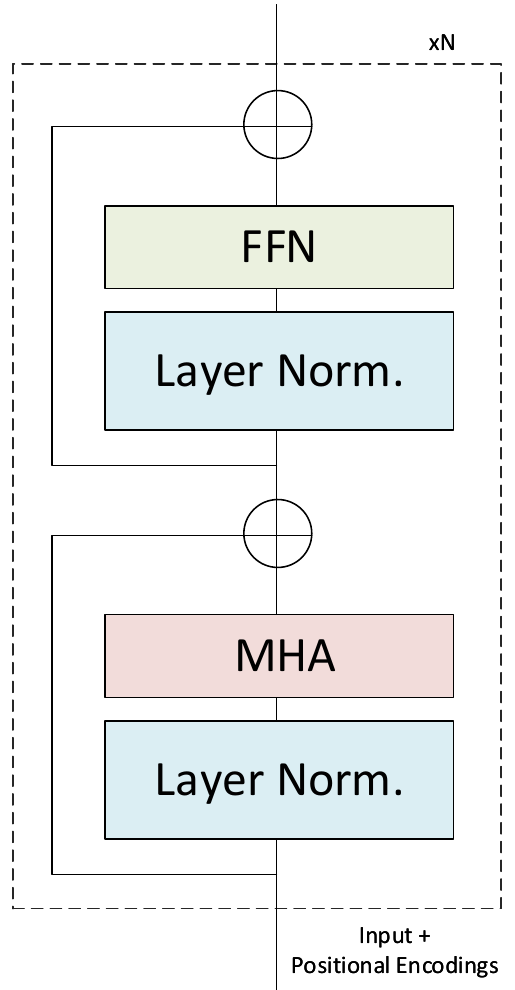}
        \label{fig:dense-trans}}
    \centering
    \subfloat[]{\includegraphics[width=0.44\columnwidth, height=65mm]{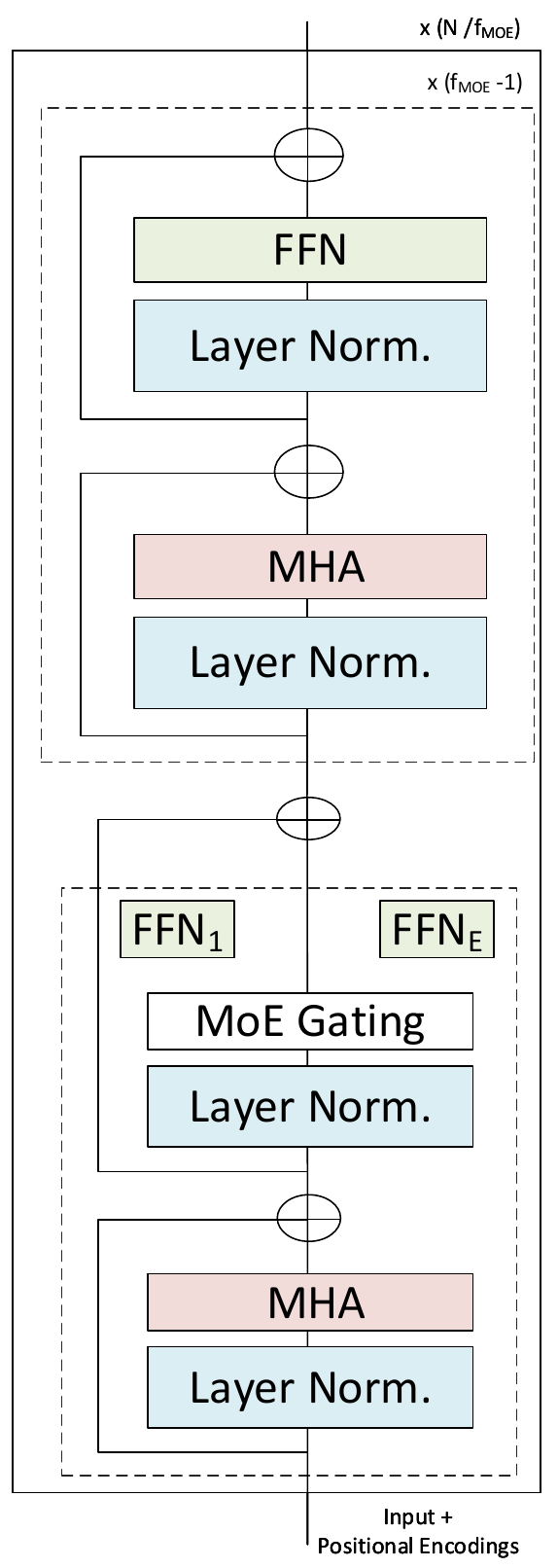}
    \label{fig:moe-trans}}
    \hspace{0.1\columnwidth}
    \caption{Illustration of Transformer encoder used, (a) Dense Transformer, (b) Mixture of Experts (MoE) Transformer.\label{fig:dense-moe}}
    \vspace{-2 mm}
\end{figure}

We have shown a comparative algorithm pipeline in Fig. \ref{fig:NLP-model}. We focused on the No Language Left Behind (NLLB-200), a translation bidirectional model supporting 200 languages, including many low-resource Indian languages, with a 600M parameter distilled INT4 lightweight version for translation across both foreign and Indic languages on resource-constrained devices.
The distilled NLLB-200 version was implemented with a 600M-parameter Transformer encoder–decoder (\figref{fig:dense-moe}) with six layers of each Pre-Norm residual connections, multi-head attention, and two-layer FFNs. Token order is decided based on positional encoding and per-language Sentence Piece tokenizers (1,000 tokens) to enable efficient many-to-many translation based on target language codes. Pretraining includes auto-encoding (DAE) for bidirectional encoding and causal language modelling (CLM) for fluency, post-finetune. The dataset involved a custom dataset, high-quality seed corpora, and generated bi-text with LASER3 embeddings for low-resource performance and balanced exposure across languages. The major resource savings originated with Mixture-of-Experts (MoE) layers that enabled the most relevant expert and processed it in parallel, combining outputs based on gated probabilities. This is crucial for enhancing parameter cost and comes at no additional per-token compute cost. Load-balancing loss penalizes skewed expert usage to avoid collapse on fixed experts, prompting even token distribution. Thus, the conditional framework scales effectively across diverse multilingual datasets and ensures robust translation even in low-resource settings.

\subsection{Hardware Architecture}

NLLB utilizes the encoder-decoder architecture (\figref{fig:dense-moe}), making it essential to design accelerators specifically for transformer workloads. Recent developments in natural language processing (NLP) involving transformer acceleration emphasize algorithm-hardware co-design to balance performance, efficiency, and resource constraints effectively. ViTCoD \cite{ViTCoD} and ViA \cite{ViA} have introduced vision transformer accelerators that prioritize sparsity and data reuse to improve throughput and reduce power consumption. EdgeBERT \cite{EdgeBERT} and QBERT \cite{QBERT-Accl} have investigated quantized NLP inference, focusing on minimizing latency and energy consumption. AccelTran \cite{AccelTran} has expanded this capability by exploiting sparsity. Platforms such as NLP-FPGA-accl \cite{NLP-FPGA-accl, HPCA'20}, OPU~\cite{OPU, Uni-OPU}, HPTA \cite{HPTA, ELSA}, and EdgeLLM \cite{EdgeLLM} demonstrate the adaptability and scalability of FPGA and ASIC SoC solutions from NLP to large language models (LLM) on heterogeneous platforms (CPU-master) with high-throughput data flow. Meanwhile, NLP-edge \cite{NLP-edge, Brainwave} showcases substantial speech and NLP inference capabilities, and LSTM-NLP \cite{LSTM-NLP} focuses on recurrent models through adaptable LSTM architectures. Consequently, we have concentrated on developing a lightweight transformer-like architecture with quantized rapid inference, assessed at both FPGA and ASIC levels, directed towards efficient NLP deployment at edge nodes. A qualitative overview of the features present in the current state-of-the-art (SoTA) Neural compute engines (NPE) is presented in \tblref{tab:SOTA-comp-features}.

\begin{table}[t]
    \caption{Qualitative comparison between SoTA AI Accelerators and features in Neural Compute Engines.}
    \label{tab:SOTA-comp-features}
    \renewcommand{\arraystretch}{1.35}
    \resizebox{\columnwidth}{!}{%
    \begin{tabular}{|l|ll|ll|l|}
        \hline
        \multicolumn{1}{|c|}{\multirow{2}{*}{\textbf{Design}}} & \multicolumn{2}{c|}{\textbf{Precision}} & \multicolumn{2}{c|}{\textbf{Design}} & \multicolumn{1}{c|}{\multirow{2}{*}{\textbf{Use-cases}}} \\ \cline{2-5}
        \multicolumn{1}{|c|}{} & \multicolumn{1}{c|}{\textbf{Datatype}} & \multicolumn{1}{c|}{\textbf{Bit-width}} & \multicolumn{1}{c|}{\textbf{Approach}} & \multicolumn{1}{c|}{\textbf{Overhead}} & \multicolumn{1}{c|}{} \\ \hline
        \textbf{JSSC'25~\cite{Occamy_JSSC}} & \multicolumn{1}{l|}{FP} & 8/16/32/64 & \multicolumn{1}{l|}{Radix-4 Booth} & \multicolumn{1}{c|}{-} & GPU Server \\ \hline
        \textbf{TCAS-I'25~\cite{Maestro}} & \multicolumn{1}{l|}{FP/BF16} & 8/16/32/64 & \multicolumn{1}{l|}{LPC-DOTP} & \multicolumn{1}{c|}{-} & AIoT \\ \hline
        \textbf{TCAD'25~\cite{AMD-MACC-TCAD'25}} & \multicolumn{1}{l|}{FP/TF32/BF16} & 4/8/16/32 & \multicolumn{1}{l|}{LUT} & \multicolumn{1}{c|}{Power} & Versal MPSoC \\ \hline
        \begin{tabular}[c]{@{}l@{}}\textbf{Edge-BERT~\cite{EdgeBERT},}\\ \textbf{Accel-Tran~\cite{AccelTran},}\\ \textbf{EdgeLLM~\cite{EdgeLLM}}\end{tabular} & \multicolumn{1}{l|}{INT/FP} & 4/8/16 & \multicolumn{1}{l|}{\begin{tabular}[c]{@{}l@{}}Grouped/Tiled\\ Matrix Compute\end{tabular}} & \begin{tabular}[c]{@{}l@{}}Resources utilization\\ (Dark-Silicon)\end{tabular} & \begin{tabular}[c]{@{}l@{}}Edge-AI (NLP)\\ Transformers\end{tabular} \\ \hline
        \textbf{HCS'24~\cite{NVIDIA-blackwell}} & \multicolumn{1}{l|}{MXFP/BF16} & 4/6/8/16 & \multicolumn{1}{l|}{Mixed-precision} & \multicolumn{1}{c|}{-} & GPU \\ \hline
        \textbf{ISCAS'25~\cite{LPRE}} & \multicolumn{1}{l|}{L. Posit} & 8/16/32 & \multicolumn{1}{l|}{Approximation} & Accuracy & Edge Compute \\ \hline
        
        \textbf{MICRO'24~\cite{AMD-XDNA}} & \multicolumn{1}{l|}{INT/BF16} & 4/8/16/32 & \multicolumn{1}{c|}{-} & Memory Bound Compute & Mobile PC (NPUs) \\ \hline
        \textbf{JSSC'23~\cite{JSSC-Samsung}} & \multicolumn{1}{l|}{INT/FP} & 4/8/16 & \multicolumn{1}{l|}{\begin{tabular}[c]{@{}l@{}}Parallel Multiplier, \\ Adder Tree Acc.\end{tabular}} & \begin{tabular}[c]{@{}l@{}}Under-utilization\\ (Dark-Silicon)\end{tabular} & Mobile SoC \\ \hline
        \begin{tabular}[c]{@{}l@{}}\textbf{ELSA, QBERT~\cite{ELSA},}\\\textbf{ViA, HPTA~\cite{ViA, HPTA}}\\ \textbf{Brainwave~\cite{Brainwave},}\end{tabular} & \multicolumn{1}{l|}{INT/FxP} & 4/8/16 & \multicolumn{1}{l|}{\begin{tabular}[c]{@{}l@{}}Flexible\\ LUT\end{tabular}} & Latency, Control & NLP (Transformers) \\ \hline
        \textbf{TCAS-II'24~\cite{FMA-TCASII'24}} & \multicolumn{1}{l|}{FP/BF16/TF32} & 16/32/64 & \multicolumn{1}{l|}{HPS, CEC} & \begin{tabular}[c]{@{}l@{}}Area (Mant. multiplication)\\ Delay (Exp. processing)\end{tabular} & \begin{tabular}[c]{@{}l@{}}High Performance\\ Computing\end{tabular} \\ \hline
        \textbf{ISCA'21~\cite{RAPID-IBM}} & \multicolumn{1}{l|}{FP/FxP} & 2/4/8/16 & \multicolumn{1}{l|}{Approximation} & Accuracy Drop & - \\ \hline
        \textbf{VLSID-26~\cite{VLSID'26}} & \multicolumn{1}{l|}{Posit/FP} & 4/8/16 & \multicolumn{1}{l|}{LPC, PEC} & Delay & XR Perception \\ \hline
        \textbf{This Work} & \multicolumn{1}{l|}{INT/FP/BF} & 4/8/16 & \multicolumn{1}{l|}{RMMEC} & Run-time adaptivity & Edge-NLP\\ \hline
    \end{tabular}}
\end{table}

\begin{figure}[!t]
    \centering
    \includegraphics[width=0.85\columnwidth]{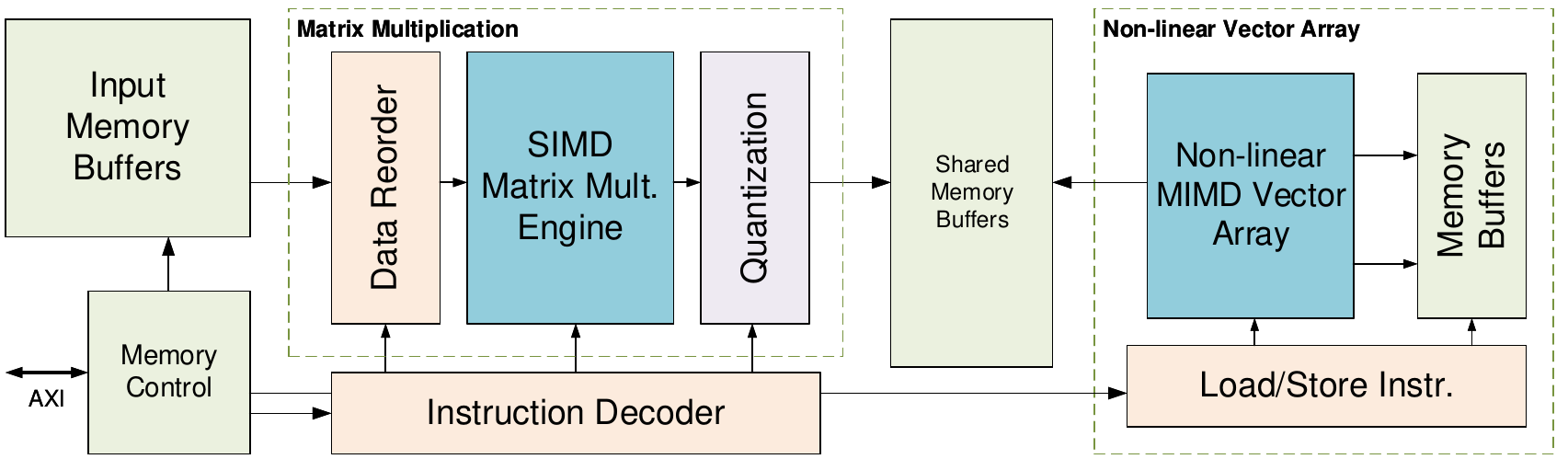}
    \caption{The detailed flow for NLPE, Memory control handles read/write with off-chip memory, NVU handles non-linear operations, while SIMD Matrix Mult. Engine handles quantized matrix multiplication.}
    \label{fig:NLPE}
\end{figure}

\begin{figure}
    \centering
    \includegraphics[width=0.85\columnwidth]{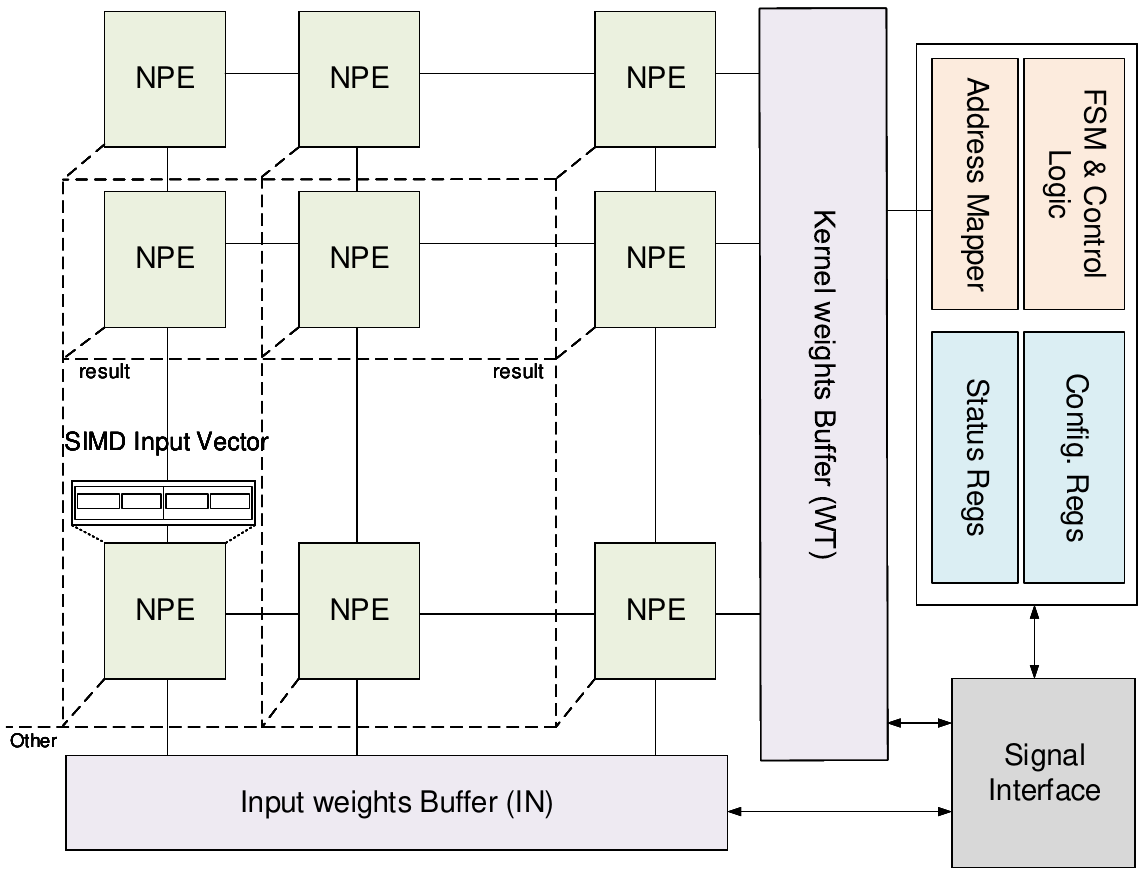}
    \caption{The detailed SIMD Matrix Mult. Engine datapath.}
    \label{fig:SIMDmatmul}
\end{figure}

\begin{figure}
    \centering
    \includegraphics[width=0.75\linewidth]{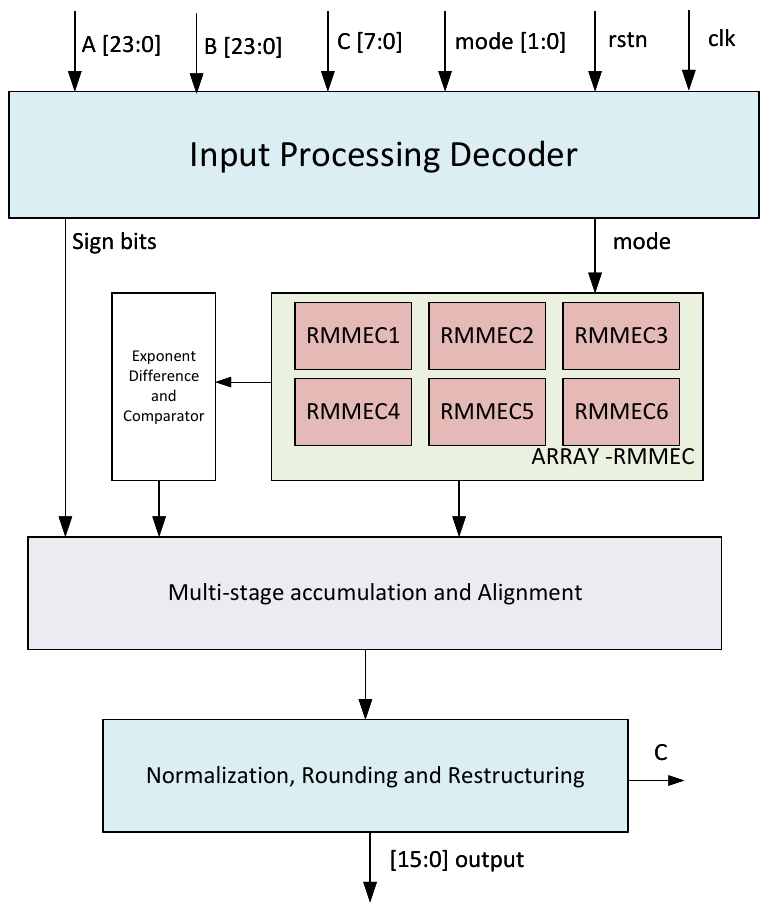}
    \caption{The datapath for SIMD multiply-accumulate unit supporting precision 6\X INT4/FP4, 3\X{FP8}, 1\X{BF16}.}
    \label{fig:MAC}
\end{figure}

\begin{figure}
    \centering
    \includegraphics[width=0.65\columnwidth]{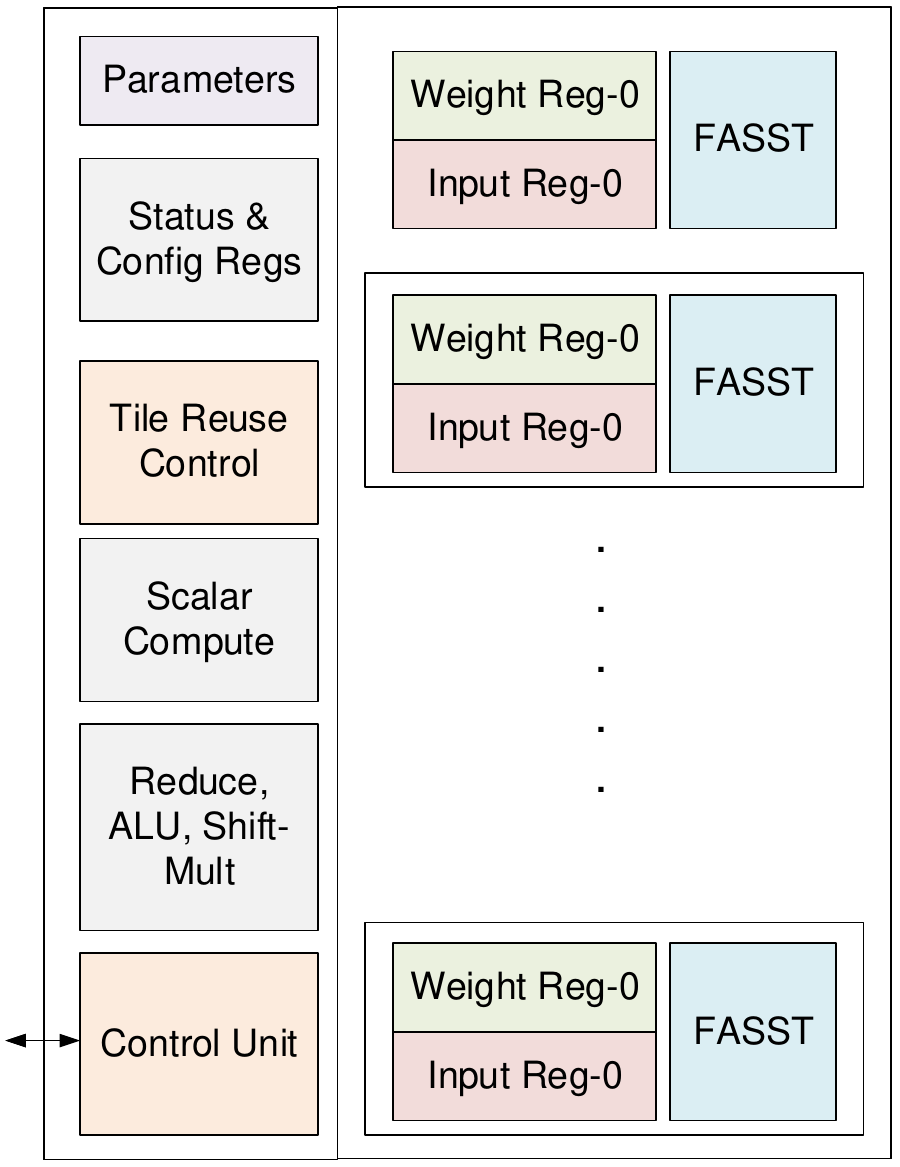}
    \caption{The detailed Non-linear MIMD Vector Array datapath.}
    \label{fig:NMV}
\end{figure}

\begin{figure}
    \centering
    \includegraphics[width=0.85\columnwidth]{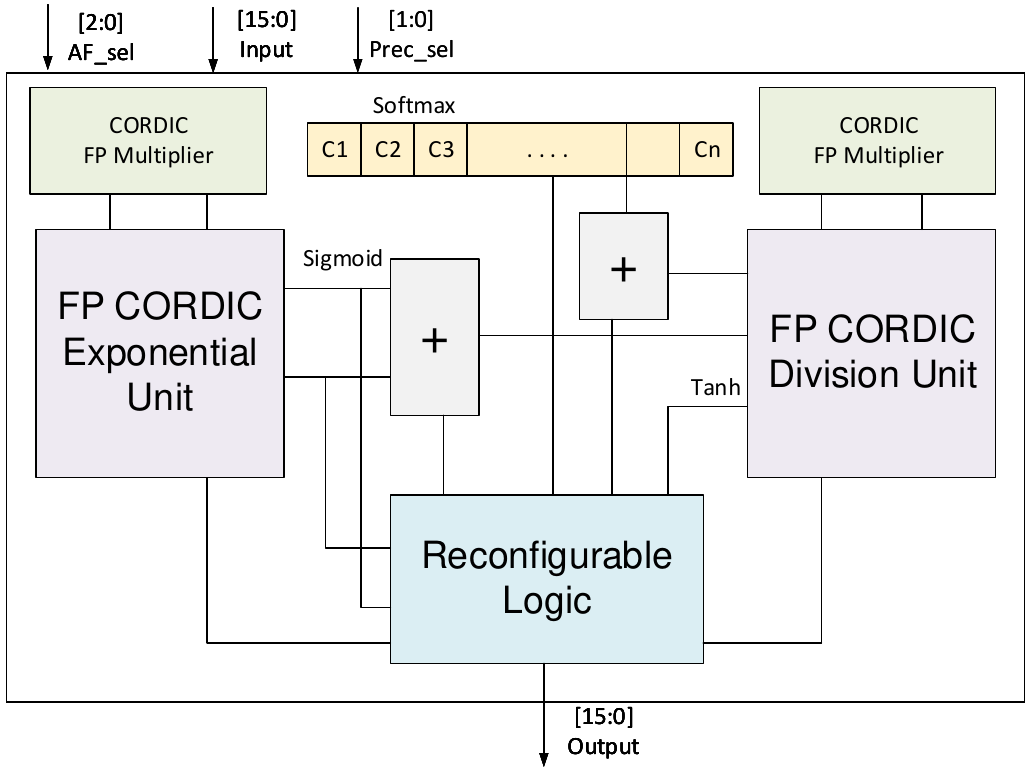}
    \caption{The detailed CORDIC-based FASST (Floating-point Activation function unit for SoftMax-sigmoid-tanh datapath.}
    \label{fig:NAF}
\end{figure}

The NLPE architecture (\figref{fig:NLPE}) is composed of modules including the Control Unit (CU), Memory Read Unit (MRU), Memory Write Unit (MWU), Matrix Multiply Engine (MME) (\figref{fig:SIMDmatmul}) based on SIMD MAC (\figref{fig:MAC}), and the Nonlinear MIMD Vector array (NMV) (\figref{fig:NMV}). The CU orchestrates execution by sending instructions to all functional units. The MRU retrieves data from external memory into the MME, while the MWU saves results back to external memory. The MME performs matrix multiplications with an array of SIMD NPEs, supported by scratchpad memory. This process involves selecting data for loading and rearranging operands from input buffers, with parallel multiplications accumulated in quantized format. The NMV manages a MIMD array of non-linear activation functions (NAFs) (\figref{fig:NAF}) with throughput supporting 2\X{FP8}/1\X{BF16} parallel operations for sigmoid, tanh, ReLU, and SoftMax functions, utilizing shared CORDIC resources~\cite{FP-CORDIC-AF}. Parallel load/store vector instructions enable simultaneous handling of multiple operands in each cycle. All units are pipelined, permitting overlaps between computation and data movement, thus reducing off-chip memory latency and optimizing concurrency by facilitating data exchange between functional units through individual memory buffers.

The parametrized systolic array based on SIMD MAC (\figref{fig:MAC}) is designed for multiple attention heads in output-stationary dataflow, where the partial results of the computation are held stationary within the PE, and the weights and embeddings are passed horizontally and vertically. The size of the systolic array is kept parameterized depending on the dimensions of the matrices, which also determines cycles for accumulation. The same systolic array hardware can be used to perform calculations for multiple attention head layers by changing the weights and embedding matrices in consecutive cycles. The MAC is divided in five stages and supports 1 \X BF16, 3 \X FP8, 6 \X FP4, and 6 \X INT4 multiply operations with one addend in a single clock cycle and output either in FP8 or BF16 which are precision supported by FASST. The novel RMMEC 4-bit blocks can be configured to operate as a multiplier or an exponent comparator by applying a mode control signal. The input processing decoder unit receives two input data values, 24-bit $A$ and $B$, one addend value $C$ (8/16-bit), and a control signal. The 24-bit input size allows for full resource utilization in SIMD mode. 

\begin{figure*}
    \centering
    \includegraphics[width=0.9\linewidth]{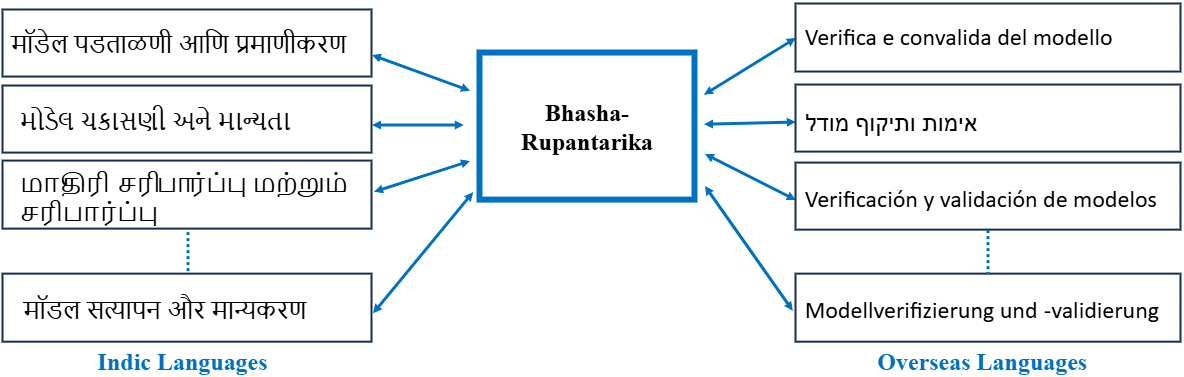}
    \caption{True outputs with Bhasha-Rupantarika (INT4) in different Indic and overseas languages, showing translation for the phrase ``Model Verification and Validation''. }
    \label{fig:output}
\end{figure*}

The sign, exponent, and mantissa bits are extracted, with the mantissa divided into 4-bit slices for input to basic multiplier blocks, while the exponent values are cut into primary blocks configured with comparators. A 2\X{3} array of 4-bit RMMEC blocks, receiving nibble inputs, generates 8-bit partial products with positional weights; a comparator assesses the maximum among the three exponents and calculates their differences for exponent normalization. For radix normalization, each exponent is adjusted by subtracting the maximum exponent value using a subtractor array, then used to shift-align the products. These aligned values are combined using Carry Select Adders (CSA) to yield the accumulated result, and the accumulator's value is recalibrated for further computations, with the exponent set to the prior maximum value. An XOR operation determines the sign of the multiplication result, obtained from the final carry out in the adder stage, with the radix point adjusted by counting leading zeros and shifting as necessary. The mantissa result is stored as 8/16-bit depending on the accumulation mode, and the exponent as 4/8-bit during the quire stage; it can be truncated, or an exception flag is set once the entire vector's dot product is complete. The upper bits, as per precision, are retained for the mantissa and recombined with the sign and exponent components.

Another crucial component is the Non-linear MIMD vector array datapath (Fig. \figref{fig:NMV}), constructed using a FASST unit based on CORDIC. This unit predominantly handles computations of various NAFs in a SIMD manner and accommodates FP8 and BF16 data precisions, with support for numerous instructions across different NAFs. It adopts an accumulative approach, synthesizing methods from \cite{FP-CORDIC-AF, Retro}. The unit employs CORDIC floating-point calculations for exponential functions and division, with reconfigurable logic aiding in operations such as swish, GeLU, selu, and SoftMax. Our implementation extends mathematical formulations from \cite{FP-CORDIC-AF}, incorporating FP8 support and hardware reuse. Previous research confined functions to GeLU/sigmoid/tanh in SRNNs, LSTMs, GRUs, Transformers, BERT, and GPT2 \cite{NLP-FPGA-accl, LSTM-NLP}, and GeLU/SoftMax in others \cite{QBERT-Accl}. Solutions like Flex-SFU and ASTRA \cite{ASTRA} provided GeLU/SiLU and SoftMax/GeLU support for BERT/GPT-2 Language modelling, while PACE \cite{Davide_Jetcas} focused on SoftMax/GeLU for Gen-AI, and ReAFM on swish, GeLU, prelu, etc., utilizing Taylor series, piecewise linear or logarithmic approximations, LUT-based methods, stochastic computation, and CORDIC-based reconfigurability. However, earlier designs did not support most NAFs with a single, reusable hardware, a gap we addressed to enhance the resource efficiency and functionality of the unified vector array. This was essential as NAF hardware can consume up to 20–25\% of the area in a commercial Google TPUv4.

\section{Performance Evaluation}

\begin{figure}
    \centering
    \includegraphics[width=0.95\linewidth]{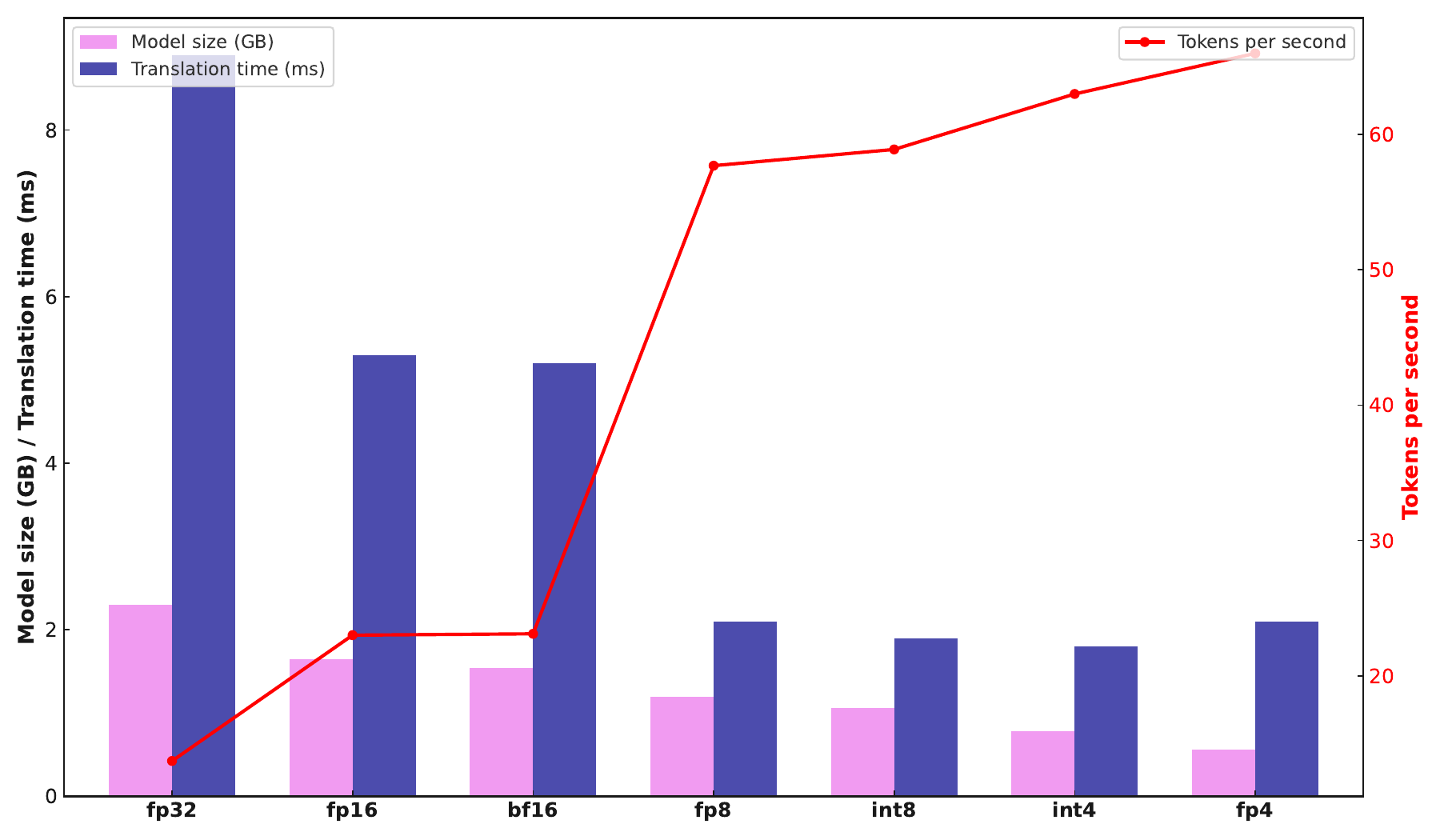}
    \caption{Performance comparison for NLLB-600M across different quantized precision.}
    \label{fig:nllb_600m_perf}
\end{figure}

The experimental setup included a software evaluation of the NLLB model, where quantized INT4/INT8 and FP4/FP8 were implemented and compared with the Baseline (FP). For resource utilization, the hardware setup employed a ZCU104 FPGA. Architectural emulation was aligned with hardware design for the benefits of co-design. Our approach utilized precise arithmetic for MAC operations and implemented an iso-functional arithmetic class for the activation layer. Post-training quantization (PTQ) was conducted using the BitsAndBytesConfig library, and performance was evaluated on 1000 queries per language, yielding satisfactory results. The setup included Python 3.0 and Qkeras 2.3 within Jupyter Notebook, operating on an NVIDIA L4 GPU. The methodology applied the QLoRA (Quantized Low-Rank Adapter) framework to fine-tune large language models (LLMs) by combining low-rank adaptation with 4-bit quantization. This approach maintains the original quantized weights while introducing small, trainable adapter modules instead of updating all model parameters. This reduces memory and computational demands and maintains performance, allowing fine-tuning of very large NLP models on consumer-grade GPUs. Additionally, 8-bit block-wise quantization was utilized. We employed the Whisper model for speech-to-text and Google Text-to-Speech (gTTS) for the text-to-speech model. Performance comparisons with various numerical precisions, with an emphasis on quantization, are illustrated in \figref{fig:nllb_600m_perf}, demonstrating significant reductions in memory and latency, with FP4 achieving a footprint of 0.56 GB and a throughput of 66 tokens/s.

\begin{table}[!t]
    \caption{FPGA Resource utilization for SoTA MAC designs}
    \label{tab:mac-fpga}
    \renewcommand{\arraystretch}{1.25}
    \resizebox{\columnwidth}{!}{%
    \begin{tabular}{|c|c|ccccc|}
        \hline
        \multirow{2}{*}{Design} & \multirow{2}{*}{Precision} & \multicolumn{5}{c}{FPGA Utilization (Virtex 707)} \\ \cline{3-7} 
         &  & \multicolumn{1}{c|}{LUTs} & \multicolumn{1}{c|}{FFs} & \multicolumn{1}{c|}{\begin{tabular}[c]{@{}c@{}}Delay \\ ($\mu$s)\end{tabular}}  & \multicolumn{1}{c|}{\begin{tabular}[c|]{@{}c@{}}Power \\ (mW)\end{tabular}} & \begin{tabular}[c]{@{}c@{}}Arith. Intensity \\ (pJ/Op)\end{tabular} \\ \hline
        \multirow{2}{*}{AMD IP (FPGA)} & INT4 & \multicolumn{1}{c|}{53} & \multicolumn{1}{c|}{28} & \multicolumn{1}{c|}{3.09} & \multicolumn{1}{c|}{3.48} & 10.75 \\ \cline{2-7} 
         & INT8 & \multicolumn{1}{c|}{130} & \multicolumn{1}{c|}{44} & \multicolumn{1}{c|}{3.816} & \multicolumn{1}{c|}{7.26} & 27.7 \\ \cline{2-7} 
         & INT16 & \multicolumn{1}{c|}{369} & \multicolumn{1}{c|}{76} & \multicolumn{1}{c|}{9.051} & \multicolumn{1}{c|}{16.9} & 153 \\ \cline{2-7} 
         & INT32 & \multicolumn{1}{c|}{1426} & \multicolumn{1}{c|}{214} & \multicolumn{1}{c|}{5.931} & \multicolumn{1}{c|}{22} & 130.4 \\ \hline
        \multirow{4}{*}{TVLSI'25~\cite{Flex-PE}} & Ad-FXP8 & \multicolumn{1}{c|}{256} & \multicolumn{1}{c|}{224} & \multicolumn{1}{c|}{5.98} & \multicolumn{1}{c|}{9.23} & 55.2 \\ \cline{2-7} 
         & Ad-FXP16 & \multicolumn{1}{c|}{427} & \multicolumn{1}{c|}{369} & \multicolumn{1}{c|}{6.5} & \multicolumn{1}{c|}{11.78} & 76.57 \\ \cline{2-7} 
         & Ad-FXP32 & \multicolumn{1}{c|}{681} & \multicolumn{1}{c|}{745} & \multicolumn{1}{c|}{7.34} & \multicolumn{1}{c|}{31} & 227.54 \\ \cline{2-7} 
         & SIMD-Pipelined 8/16/32 & \multicolumn{1}{c|}{897} & \multicolumn{1}{c|}{1231} & \multicolumn{1}{c|}{11.7} & \multicolumn{1}{c|}{59.4} & 694 \\ \hline
        \multirow{4}{*}{ISCAS'25~\cite{LPRE}} & Posit8 & \multicolumn{1}{c|}{467} & \multicolumn{1}{c|}{175} & \multicolumn{1}{c|}{2.68} & \multicolumn{1}{c|}{68} & 182.24 \\ \cline{2-7} 
         & Posit16 & \multicolumn{1}{c|}{2083} & \multicolumn{1}{c|}{528} & \multicolumn{1}{c|}{4.35} & \multicolumn{1}{c|}{189} & 822.15 \\ \cline{2-7} 
         & Posit32 & \multicolumn{1}{c|}{6813} & \multicolumn{1}{c|}{806} & \multicolumn{1}{c|}{8} & \multicolumn{1}{c|}{347} & 2776 \\ \cline{2-7} 
         & SIMD-L.Posit & \multicolumn{1}{c|}{4613} & \multicolumn{1}{c|}{2078} & \multicolumn{1}{c|}{6.2} & \multicolumn{1}{c|}{276} & 426 \\ \hline
        TCAS-II'24~\cite{RPE-TCASII'24} & SIMD-INT4/FP8/16/32 & \multicolumn{1}{c|}{8054} & \multicolumn{1}{c|}{1718} & \multicolumn{1}{c|}{4.62} & \multicolumn{1}{c|}{296} & 152 \\ \hline
        TCAS-II'24~\cite{RPE-TCASII'24} & SIMD-FP16/32/64 & \multicolumn{1}{c|}{8065} & \multicolumn{1}{c|}{1072} & \multicolumn{1}{c|}{5.56} & \multicolumn{1}{c|}{378} & 543 \\ \hline
        \multirow{4}{*}{Access'24~\cite{QuantMAC}} 
        & Q-4b & \multicolumn{1}{c|}{24} & \multicolumn{1}{c|}{16} & \multicolumn{1}{c|}{0.98} & \multicolumn{1}{c|}{2.2} & 2.16 \\ \cline{2-7} 
        & Q-8b & \multicolumn{1}{c|}{52} & \multicolumn{1}{c|}{88} & \multicolumn{1}{c|}{1.57} & \multicolumn{1}{c|}{6.36} & 10 \\ \cline{2-7} 
         & Q-16b & \multicolumn{1}{c|}{106} & \multicolumn{1}{c|}{168} & \multicolumn{1}{c|}{2.2} & \multicolumn{1}{c|}{11.77} & 26 \\ \cline{2-7} 
         &  \multicolumn{1}{c|}{\begin{tabular}[c]{@{}c@{}}SIMD Quant-MAC\\ (FXP8/16/32)\end{tabular}} & \multicolumn{1}{c|}{1502} & \multicolumn{1}{c|}{2418} & \multicolumn{1}{c|}{40.32} & \multicolumn{1}{c|}{21.38} & 108 \\ \hline
        TCAS-I'22~\cite{VLSID'26} & FXP8 & \multicolumn{1}{c|}{238} & \multicolumn{1}{c|}{32} & \multicolumn{1}{c|}{2.75} & \multicolumn{1}{c|}{2.8} & 7.6 \\ \hline
        CORDIC & FXP-4 & \multicolumn{1}{c|}{35} & \multicolumn{1}{c|}{58} & \multicolumn{1}{c|}{1.406} & \multicolumn{1}{c|}{4.36} & 6.13 \\ \cline{2-7} 
         & FXP-8 & \multicolumn{1}{c|}{54} & \multicolumn{1}{c|}{88} & \multicolumn{1}{c|}{1.518} & \multicolumn{1}{c|}{6.6} & 10 \\ \cline{2-7} 
         & FXP-16 & \multicolumn{1}{c|}{95} & \multicolumn{1}{c|}{162} & \multicolumn{1}{c|}{2.124} & \multicolumn{1}{c|}{12.21} & 25.9 \\ \hline
        Proposed & INT4/FP4-/8/BF16 & \multicolumn{1}{c|}{425} & \multicolumn{1}{c|}{268} & \multicolumn{1}{c|}{3.2} & \multicolumn{1}{c|}{19.56} & 10.43\\\hline
    \end{tabular}}
\end{table}

\begin{figure}
    \centering
    \includegraphics[width=1\columnwidth]{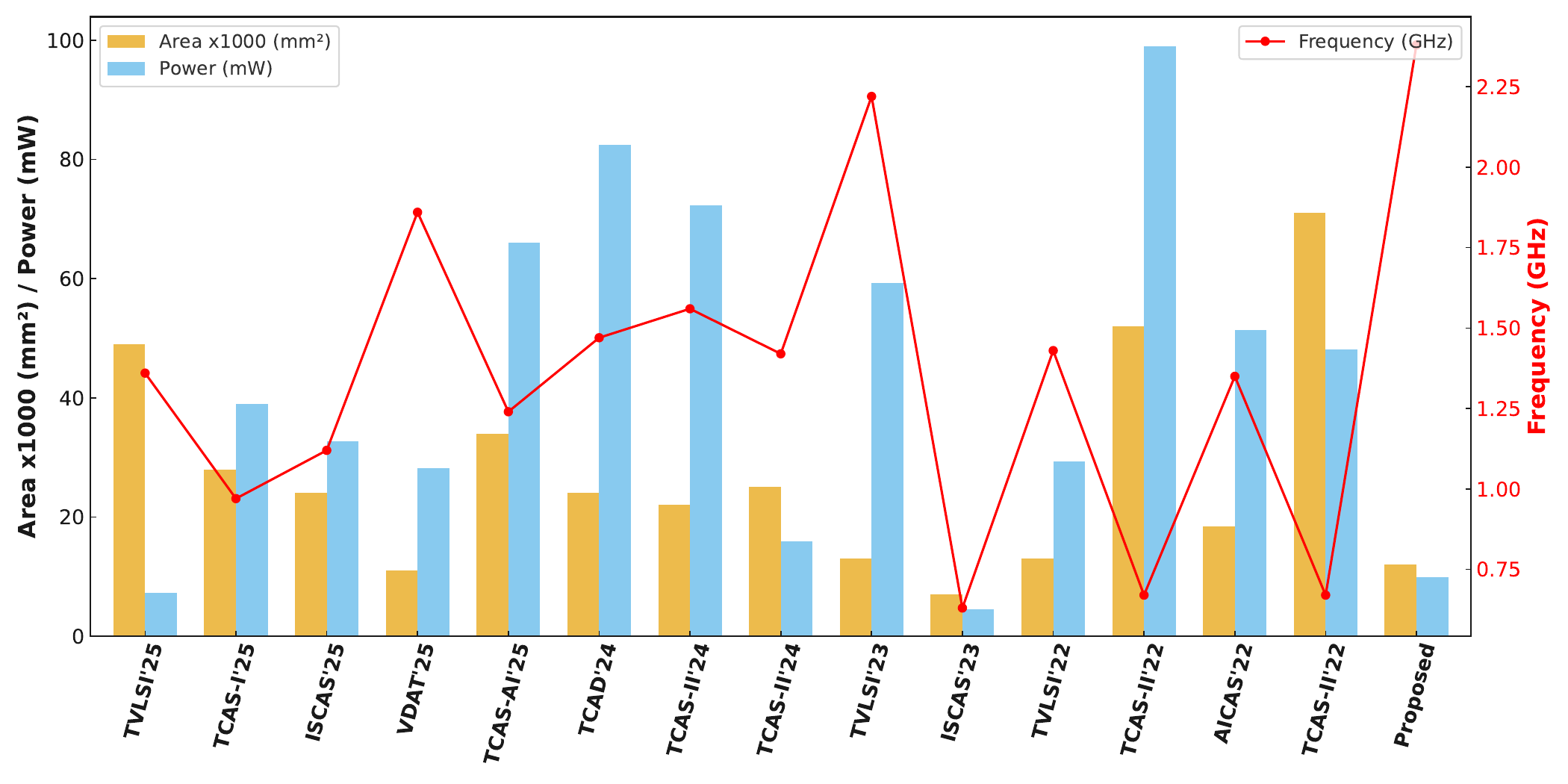}
    \caption{ASIC performance metrics, comparison with SoTA MAC units CMOS 28nm, comparison data from \cite{ReNPU}.}
    \label{fig:asic_mac}
\end{figure}

\begin{figure}[!t]
    \centering
    \includegraphics[width=0.975\columnwidth]{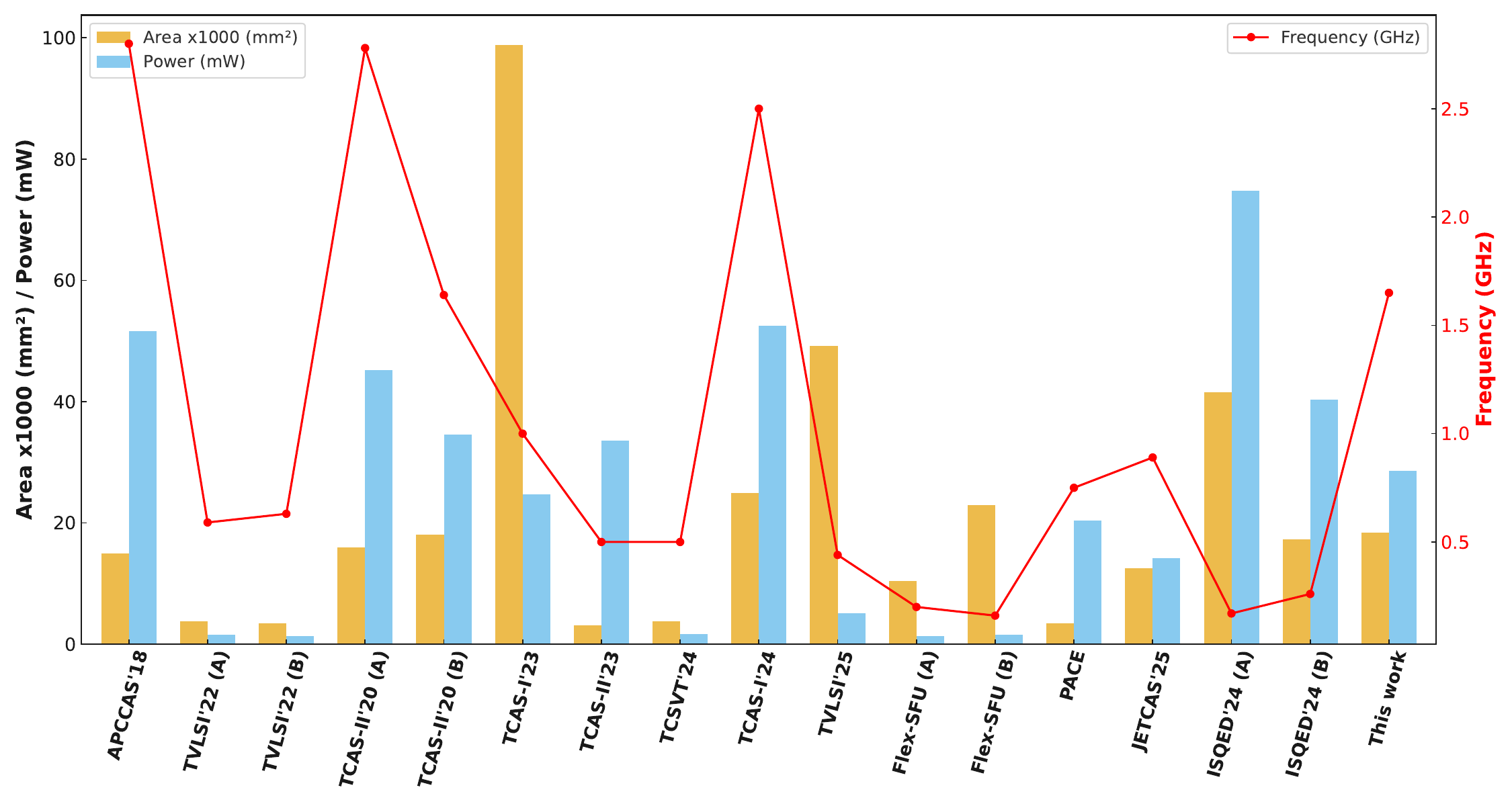}
    \caption{ASIC performance metrics, comparison with SoTA NAF units CMOS 28nm, comparison with \cite{Flex-PE, FP-CORDIC-AF, TVLSI_SoftMax'23, ASTRA, QForce-RL, Davide_Jetcas, Flex-SFU, Retro, TEAS}.}
    \label{fig:asic_naf}
\end{figure}

We designed the proposed NLPE using System Verilog for its components and verified its functionality with the Questa-Sim Simulator, making a fair comparison with an iso-functional Python emulator. In addition, FPGA synthesis was conducted at the level of fundamental modules for MAC and NAF designs. We compared these designs with SoTA systems as shown in \tblref{tab:mac-fpga} and \tblref{tab:fpga_act}, employing the AMD Vivado Design Suite. The MAC and NAF units were evaluated against individual precision units and SIMD compute units to demonstrate the efficacy of our approach. As indicated in \tblref{tab:mac-fpga}, our unit significantly lowers LUT resource usage by 90\% compared to \cite{LPRE} and 94.7\% compared to \cite{RPE-TCASII'24}; additionally, power consumption is reduced by a factor of 3$\times$ relative to \cite{Flex-PE}, and delay is reduced by up to 1.94$\times$ and 3.66$\times$ compared to \cite{LPRE} and \cite{Flex-PE}, respectively. Likewise, the NAF unit was also compared favorably against SoTA works, meeting our resource-efficiency goals for accelerating NLLB, Furthermore, the design was synthesized using Cadence Innovus in a commercial CMOS 28nm technology, and we reported pertinent performance metrics, comparing against SoTA MAC and NAF units as depicted in \figref{fig:asic_mac} and \figref{fig:asic_naf}, respectively. Our MAC units exhibit notably increased operational frequency and reductions in area and power consumption, leading to a conclusion of enhanced energy efficiency and a compact solution. The NAF was optimized, providing a superior balance between frequency and area-power consumption, as workload characterization indicates that the NAF accounts for up to 60\% of the operations in the overall NLLB.

\begin{table}
    \caption{FPGA Resource comparison for NAF functions used in NLP accelerators.}
    \label{tab:fpga_act}
    \renewcommand{\arraystretch}{1.25}
    \resizebox{\columnwidth}{!}{%
    \begin{tabular}{|c|c|c|c|c|c|c|}
        \hline
        Design & Function & Precision & Op. Freq. (MHz) & LUTs & FFs & Energy (pJ) \\ \hline
        APCCAS'18 & SoftMax & 8/16-bit & 436 & 2564 & 2794 & 1723 \\ \hline
        \multirow{4}{*}{TVLSI'23~\cite{TVLSI_SoftMax'23}} & \multirow{4}{*}{SoftMax} & DHS8 & 251.3 & 1746 & 1386 & 740 \\ \cline{3-7} 
         &  & IDHS8 & 264 & 1604 & 1354 & 698 \\ \cline{3-7} 
         &  & DHS12 & 255.8 & 1589 & 1235 & 696 \\ \cline{3-7} 
         &  & IDHS12 & 267 & 1450 & 1220 & 664 \\ \hline
        \multirow{2}{*}{TVLSI'25~\cite{ASTRA}} & \multirow{2}{*}{SoftMax/GeLU} & \multirow{2}{*}{INT16} & 435 & 1603 & 704 & 2466 \\ \cline{4-7} 
         &  &  & 500 & 1446 & 652 & 1808 \\ \hline
        TVLSI'25~\cite{Flex-PE} & SST*, relu & FxP8/16/32 & 85 & 897 & 1231 & 696 \\ \hline
        TCAS-II'23~\cite{TVLSI_SoftMax'23} & SST*, relu & 12-bit & 284 & 367 & 298 & - \\ \hline
        \multirow{6}{*}{ISQED'24~\cite{FP-CORDIC-AF}} & SoftMax & FP32 & 10.88 & 3217 & - & 11033 \\ \cline{2-7} 
         & sigmoid & FP32 & 8.27 & 5101 & - & 13189 \\ \cline{2-7} 
         & tanh & FP32 & 17.66 & 4298 & - & 7358 \\ \cline{2-7} 
         & SoftMax & BF16 & 22.18 & 1263 & - & 3472 \\ \cline{2-7} 
         & sigmoid & BF16 & 22.5 & 1856 & - & 3694 \\ \cline{2-7} 
         & tanh & BF16 & 26.4 & 1513 & - & 3100 \\ \hline
        This work & SST*, relu & FP8/BF16 & 192 & 1434 & 1208 & 987 \\ \hline
    \end{tabular}}
\end{table}

\begin{table*}[!t]
    \caption{FPGA resource utilization, comparison with prior NLP accelerator designs.}
    \label{tab:fpga-accl}
    \renewcommand{\arraystretch}{1.125}
    \resizebox{\textwidth}{!}{%
    \begin{tabular}{|c|c|c|c|c|c|c|c|c|c|c|}
        \hline
        Design & Model & FPGA & \begin{tabular}[c]{@{}c@{}}Op. Freq. \\ (MHz)\end{tabular} & Precision & \multicolumn{1}{l|}{Power (W)} & \multicolumn{1}{l|}{LUT (K)} & \multicolumn{1}{l|}{FF (K)} & \multicolumn{1}{l|}{DSP (K)} & BRAM (K) & \begin{tabular}[c]{@{}c@{}}Throughput \\ GOPS\end{tabular}\\ \hline
        NPE~\cite{OPU} & BERT & Zynq Z-7100 & 200 & 16-bit & 20 & 156 & 261 & 2.02 & 0.53 &  \\ \hline
        ISQED'21 & Transformer & Alveo U200 & - & - & 25 & 472 & 378 & 2.34 & - & 34 \\ \hline
        TECS'21~\cite{Algo-HW} & Multi-30K & ZCU102 & 150 & INT8 & 18 & 252 & 161 & 2.52 & 0.912 & 1870 \\ \hline
        TCAS-I'22~\cite{ShortcutFusion} & Yolov3 & KCU15 & 200 & 8-bit & 4.6 & 213.3 & 352 & 2.24 & - &  \\ \hline
        TPDS'22~\cite{TPDS} & NMT & VCU118 & 100 & FP16 & 30 & 556 & 520 & 4.84 & - & 22 \\ \hline
        TCAD'23~\cite{ViA} & Swin-T & Alveo U50 & 300 & FP16 & 39 & 258 & 257 & 2.42 & 1 & 309.6 \\ \hline
        HPTA~\cite{HPTA} & Swin-T & ZCU102 & 200 & 8-bit & 20 & 210 & 368 & 2.31 & 0.35 & 148.8 \\ \hline
        Q-BERT~\cite{QBERT-Accl} & MNLI & ZCU102 & 214 & 4/8-bit & 9.8 & 274 & 548 & 2.52 & 1.83 & - \\ \hline
        This Work & NLLB & ZCU104 & 250 & \begin{tabular}[c]{@{}c@{}}INT4/FP4 \\ FP8/BF16\end{tabular} & 8.12 & 79.4 & 97.24 & 1.4 & - & 684.48 \\ \hline
    \end{tabular}}
\end{table*}
For comparison, we implemented the NLPE architecture with System Verilog RTL on ZCU104 MPSoC, and reported the resources in  \tblref{tab:fpga-accl}. Based on available FPGAs, we believe ZCU is a fair comparison of FPGA resources, and our design showcases a significant resource reduction, approximately 1.96$\times$ reduction in LUTs and 1.65$\times$ reduction in FFs compared to the best of SoTA. It was worth noticing improved throughput by a factor of 2.2 $\times$ compared to \cite{OPU} and 4.6 $\times$ compared to \cite{HPTA}. The results demonstrate a lightweight language translation tool for remote areas with quick FPGA deployment.

The importance of algorithm-hardware codesign in the NLP edge deployment arena is centered on low-precision quantization, which results in reduced model sizes and accelerated inference, ideal for real-time translation on IoT platforms, with tightly constrained hardware resources (Fig. \ref{fig:output}). Additionally, there is a focus on energy-per-token efficiency for sustainable deployment. We hypothesise that examining scalability for larger models could enhance high-performance cloud implementations like chatbots, offering substantial hardware optimisation. A comprehensive error analysis would highlight the exact translation performance in terms of grammatical errors, semantic shifts, and contextual inconsistencies, particularly with local dialects. We contend that employing mixed-precision deployment strategies is vital for maintaining a balance between accuracy and resource constraints.

\section{Conclusion}

In this study, we evaluated the sub-octet quantization for multilingual translation, demonstrating significant improvements in memory usage, latency, and throughput, without compromising translation capability. Specifically, the FP4 model's size was reduced to 0.56 GB, achieving a 4.1\X reduction compared to FP32, which resulted in a throughput of 66 tokens per second, suitable for real-time processing on IoT platforms. The core components of the accelerator, MAC and NAF, outperformed previous work at both FPGA and ASIC levels. FPGA deployment confirmed the algorithm's benefits, showing a 1.96\X reduction in LUTs, a 1.65\X reduction, and a throughput increase by a factor of 2.2\X compared to OPU and 4.6\X compared to HPTA at an operating frequency of 250 MHz. Overall, the findings establish Bhasha-Rupantarika as an effective solution for multilingual translation systems.

\bibliographystyle{ieeetr}
\bibliography{bib}

\begin{thebibliography}{10}

\bibitem{TPDS}
Q.~Li, X.~Zhang, J.~Xiong, W.-M. Hwu, and D.~Chen, ``{Efficient Methods for Mapping Neural Machine Translator on FPGAs},'' {\em IEEE Transactions on Parallel and Distributed Systems}, vol.~32, pp.~1866--1877, July 2021.

\bibitem{Algo-HW}
X.~Zhang, Y.~Wu, P.~Zhou, X.~Tang, and J.~Hu, ``{Algorithm-hardware Co-design of Attention Mechanism on FPGA Devices},'' {\em ACM Trans. Embed. Comput. Syst.}, vol.~20, Sept. 2021.

\bibitem{Raj-survey}
R.~Dabre, C.~Chu, and A.~Kunchukuttan, ``{A Survey of Multilingual Neural Machine Translation},'' {\em ACM Comput. Surv.}, vol.~53, Sept. 2020.

\bibitem{Raj2_acm}
A.~Joshi, R.~Dabre, D.~Kanojia, Z.~Li, H.~Zhan, G.~Haffari, and D.~Dippold, ``{Natural Language Processing for Dialects of a Language: A Survey},'' {\em ACM Comput. Surv.}, vol.~57, Feb. 2025.

\bibitem{Raj_acm2}
H.~Song, R.~Dabre, C.~Chu, S.~Kurohashi, and E.~Sumita, ``{SelfSeg: A Self-supervised Sub-word Segmentation Method for Neural Machine Translation},'' {\em ACM Trans. Asian Low-Resour. Lang. Inf. Process.}, vol.~22, Aug. 2023.

\bibitem{Survey}
C.~Guo, F.~Cheng, Z.~Du, {\em et~al.}, ``{A Survey: Collaborative Hardware and Software Design in the Era of Large Language Models},'' {\em IEEE Circuits and Systems Magazine}, vol.~25, pp.~35--57, Feb. 2025.

\bibitem{Indic-ST}
N.~Sethiya, S.~Nair, P.~Walia, and C.~Maurya, ``{Indic-ST: A Large-Scale Multilingual Corpus for Low-Resource Speech-to-Text Translation},'' {\em ACM Trans. Asian Low-Resour. Lang. Inf. Process.}, vol.~24, June 2025.

\bibitem{OPU}
Y.~Yu, C.~Wu, T.~Zhao, K.~Wang, and L.~He, ``{OPU: An FPGA-Based Overlay Processor for Convolutional Neural Networks},'' {\em IEEE Transactions on Very Large Scale Integration (VLSI) Systems}, vol.~28, pp.~35--47, Jan. 2020.

\bibitem{LSTM-NLP}
E.~Azari and S.~Vrudhula, ``{An Energy-Efficient Reconfigurable LSTM Accelerator for Natural Language Processing},'' in {\em 2019 IEEE International Conference on Big Data (Big Data)}, pp.~4450--4459, 2019.

\bibitem{bhasaanuvaad}
S.~Jain, A.~Sankar, D.~Choudhary, D.~Suman, N.~Narasimhan, M.~S. U.~R. Khan, A.~Kunchukuttan, M.~M. Khapra, and R.~Dabre, ``{Bhasaanuvaad: A speech translation dataset for 13 indian languages},'' {\em arXiv preprint arXiv:2411.04699}, Nov. 2024.

\bibitem{bhashini}
G.~K. Kumar, S.~Praveen, P.~Kumar, M.~M. Khapra, and K.~Nandakumar, ``{Towards building text-to-speech systems for the next billion users},'' in {\em Icassp 2023-2023 ieee international conference on acoustics, speech and signal processing (icassp)}, pp.~1--5, IEEE, May 2023.

\bibitem{NLLB}
``{Scaling neural machine translation to 200 languages},'' {\em Nature}, vol.~630, pp.~841--846, June 2024.

\bibitem{whisper}
A.~Radford, J.~W. Kim, T.~Xu, G.~Brockman, C.~McLeavey, and I.~Sutskever, ``{Robust speech recognition via large-scale weak supervision},'' in {\em International conference on machine learning}, pp.~28492--28518, PMLR, July 2023.

\bibitem{ViTCoD}
H.~You, Z.~Sun, H.~Shi, Z.~Yu, Y.~Zhao, Y.~Zhang, C.~Li, B.~Li, and Y.~Lin, ``{ViTCoD: Vision Transformer Acceleration via Dedicated Algorithm and Accelerator Co-Design},'' in {\em IEEE International Symposium on High-Performance Computer Architecture (HPCA)}, pp.~273--286, 2023.

\bibitem{ViA}
T.~Wang, L.~Gong, C.~Wang, Y.~Yang, Y.~Gao, X.~Zhou, and H.~Chen, ``{ViA: A Novel Vision-Transformer Accelerator Based on FPGA},'' {\em IEEE Transactions on Computer-Aided Design of Integrated Circuits and Systems}, vol.~41, pp.~4088--4099, Nov. 2022.

\bibitem{EdgeBERT}
T.~Tambe, C.~Hooper, L.~Pentecost, {\em et~al.}, ``{EdgeBERT: Sentence-Level Energy Optimizations for Latency-Aware Multi-Task NLP Inference},'' in {\em MICRO-54: 54th Annual IEEE/ACM International Symposium on Microarchitecture}, MICRO '21, (New York, NY, USA), p.~830–844, Association for Computing Machinery, 2021.

\bibitem{QBERT-Accl}
Z.~Liu, G.~Li, and J.~Cheng, ``Hardware acceleration of fully quantized bert for efficient natural language processing,'' {\em 2021 Design, Automation \& Test in Europe Conference \& Exhibition (DATE)}, pp.~513--516, 2021.

\bibitem{AccelTran}
S.~Tuli and N.~K. Jha, ``{AccelTran: A Sparsity-Aware Accelerator for Dynamic Inference With Transformers},'' {\em IEEE Transactions on Computer-Aided Design of Integrated Circuits and Systems}, vol.~42, pp.~4038--4051, Nov. 2023.

\bibitem{NLP-FPGA-accl}
S.~Hur, S.~Na, D.~Kwon, J.~Kim, A.~Boutros, E.~Nurvitadhi, and J.~Kim, ``{A Fast and Flexible FPGA-based Accelerator for Natural Language Processing Neural Networks},'' {\em ACM Trans. Archit. Code Optim.}, vol.~20, Feb. 2023.

\bibitem{HPCA'20}
T.~J. Ham, S.~J. Jung, {\em et~al.}, ``{A\textsuperscript{3}: Accelerating Attention Mechanisms in Neural Networks with Approximation},'' in {\em IEEE International Symposium on High Performance Computer Architecture (HPCA)}, pp.~328--341, 2020.

\bibitem{Uni-OPU}
Y.~Yu, T.~Zhao, M.~Wang, K.~Wang, and L.~He, ``{Uni-OPU: An FPGA-Based Uniform Accelerator for Convolutional and Transposed Convolutional Networks},'' {\em IEEE Transactions on Very Large Scale Integration (VLSI) Systems}, vol.~28, pp.~1545--1556, July 2020.

\bibitem{HPTA}
Y.~Han and Q.~Liu, ``{HPTA: A High Performance Transformer Accelerator Based on FPGA},'' in {\em 2023 33rd International Conference on Field-Programmable Logic and Applications (FPL)}, pp.~27--33, 2023.

\bibitem{ELSA}
T.~J. Ham, Y.~Lee, {\em et~al.}, ``{ELSA: hardware-software co-design for efficient, lightweight self-attention mechanism in neural networks},'' in {\em Proceedings of the 48th Annual International Symposium on Computer Architecture}, ISCA '21, p.~692–705, IEEE Press, 2021.

\bibitem{EdgeLLM}
M.~Huang, A.~Shen, K.~Li, {\em et~al.}, ``{EdgeLLM: A Highly Efficient CPU-FPGA Heterogeneous Edge Accelerator for Large Language Models},'' {\em IEEE Transactions on Circuits and Systems I: Regular Papers}, vol.~72, pp.~3352--3365, July 2025.

\bibitem{NLP-edge}
T.~Tambe, E.-Y. Yang, G.~G. Ko, Y.~Chai, {\em et~al.}, ``{A 16-nm SoC for Noise-Robust Speech and NLP Edge AI Inference With Bayesian Sound Source Separation and Attention-Based DNNs},'' {\em IEEE Journal of Solid-State Circuits}, vol.~58, pp.~569--581, Feb. 2023.

\bibitem{Brainwave}
J.~Fowers, K.~Ovtcharov, {\em et~al.}, ``{A configurable cloud-scale DNN processor for real-time AI},'' in {\em Proceedings of the 45th Annual International Symposium on Computer Architecture}, ISCA '18, p.~1–14, IEEE Press, 2018.

\bibitem{Occamy_JSSC}
P.~Scheffler, T.~Benz, {\em et~al.}, ``{Occamy: A 432-Core Dual-Chiplet Dual-HBM2E 768-DP-GFLOP/s RISC-V System for 8-to-64-bit Dense and Sparse Computing in 12-nm FinFET},'' {\em {IEEE Journal of Solid-State Circuits}}, vol.~60, Apr. 2025.

\bibitem{Maestro}
M.~Sinigaglia {\em et~al.}, ``{Maestro: A 302 GFLOPS/W and 19.8 GFLOPS RISC-V Vector-Tensor Architecture for Wearable Ultrasound Edge Computing},'' {\em IEEE Trans. on Circuits and Syst.- I}, pp.~1--15, 2025.

\bibitem{AMD-MACC-TCAD'25}
H.~J. Damsgaard, K.~J. HoBfeld, and J.~Nurmi, ``{Parallel Accurate Minifloat MACCs for NN Inference on Versal FPGAs},'' {\em IEEE Trans. Comp.-Aided Des. Integ. Cir. Syst.}, vol.~44, pp.~2181--2194, June 2025.

\bibitem{NVIDIA-blackwell}
A.~Tirumala and R.~Wong, ``{NVIDIA Blackwell Platform: Advancing Generative AI and Accelerated Computing},'' in {\em IEEE Hot Chips Symposium (HCS)}, vol.~36, pp.~1--33, 2024.

\bibitem{LPRE}
O.~Kokane, M.~Lokhande, G.~Raut, A.~Teman, and S.~K. Vishvakarma, ``{LPRE: Logarithmic Posit-enabled Reconfigurable edge-AI Engine},'' in {\em 2025 IEEE International Symposium on Circuits and Systems (ISCAS)}, pp.~1--5, May 2025.

\bibitem{AMD-XDNA}
A.~Rico, S.~Pareek, {\em et~al.}, ``{AMD XDNA NPU in Ryzen AI Processors},'' {\em IEEE Micro}, vol.~44, pp.~73--82, Nov. 2024.

\bibitem{JSSC-Samsung}
J.-S. Park, C.~Park, {\em et~al.}, ``{A Multi-Mode 8k-MAC HW-Utilization-Aware Neural Processing Unit With a Unified Multi-Precision Datapath in 4-nm Flagship Mobile SoC},'' {\em IEEE Journal of Solid-State Circuits}, vol.~58, pp.~189--202, Jan. 2023.

\bibitem{FMA-TCASII'24}
H.~Tan, J.~Zhang, X.~He, L.~Huang, Y.~Wang, and L.~Xiao, ``{A Low-Cost Floating-Point FMA Unit Supporting Package Operations for HPC-AI Applications},'' {\em IEEE Trans. on Circuits and Systems II: Express Briefs}, vol.~71, pp.~3488--3492, July 2024.

\bibitem{RAPID-IBM}
S.~Venkataramani, V.~Srinivasan, {\em et~al.}, ``{RaPiD: AI Accelerator for Ultra-low Precision Training and Inference},'' {\em ACM/IEEE 48th Annual International Symposium on Computer Architecture}, pp.~153--166, 2021.

\bibitem{VLSID'26}
T.~Chaudhari, T.~Dewangan, M.~Lokhande, S.~K. Vishvakarma, {\em et~al.}, ``{XR-NPE: High-Throughput Mixed-precision SIMD Neural Processing Engine for Extended Reality Perception Workloads},'' {\em arXiv preprint arXiv:2508.13049}, 2025.

\bibitem{FP-CORDIC-AF}
M.~Basavaraju, V.~Rayapati, and M.~Rao, ``{Exploring Hardware Activation Function Design: CORDIC Architecture in Diverse Floating Formats},'' in {\em 2024 25th International Symposium on Quality Electronic Design (ISQED)}, pp.~1--8, 2024.

\bibitem{Retro}
O.~Kokane, G.~Raut, S.~Ullah, M.~Lokhande, A.~Teman, A.~Kumar, and S.~K. Vishvakarma, ``{Retrospective: A CORDIC Based Configurable Activation Function for NN Applications},'' in {\em 2025 IEEE Computer Society Annual Symposium on VLSI (ISVLSI)}, vol.~1, pp.~1--6, 2025.

\bibitem{ASTRA}
H.~Shao and Z.~Wang, ``{ASTRA: Reconfigurable Training Architecture Design for Nonlinear Softmax and Activation Functions in Transformers},'' {\em IEEE Transactions on Very Large Scale Integration (VLSI) Systems}, vol.~33, pp.~2054--2058, July 2025.

\bibitem{Davide_Jetcas}
A.~Belano, Y.~Tortorella, A.~Garofalo, {\em et~al.}, ``{A Flexible Template for Edge Generative AI With High-Accuracy Accelerated Softmax and GELU},'' {\em IEEE Journal on Emerging and Selected Topics in Circuits and Systems}, vol.~15, pp.~200--216, June 2025.

\bibitem{Flex-PE}
M.~Lokhande, G.~Raut, and S.~K. Vishvakarma, ``{Flex-PE: Flexible and SIMD Multiprecision Processing Element for AI Workloads},'' {\em IEEE Trans. {VLSI} Syst.}, vol.~33, pp.~1610--1623, June 2025.

\bibitem{RPE-TCASII'24}
B.~Li, K.~Li, {\em et~al.}, ``{A Reconfigurable Processing Element for Multiple-Precision Floating/Fixed-Point HPC},'' {\em IEEE Trans. Circuits Syst. II}, vol.~71, pp.~1401--1405, Mar. 2024.

\bibitem{QuantMAC}
N.~Ashar, G.~Raut, V.~Trivedi, S.~K. Vishvakarma, and A.~Kumar, ``{QuantMAC: Enhancing Hardware Performance in DNNs With Quantize Enabled Multiply-Accumulate Unit},'' {\em IEEE Access}, vol.~12, pp.~43600--43614, 2024.

\bibitem{ReNPU}
M.~Lokhande, S.~J. Chand, A.~Jain, S.~Kumar, and S.~K. Vishvakarma, ``{ReNPU: A Resource-efficient Multi-Mode Neural Processing Unit with Unified Multi-Precision Datapath for Mobile AI Workloads},'' {\em Authorea Preprints}, 2025.

\bibitem{TVLSI_SoftMax'23}
X.~Wu, S.~Liang, M.~Wang, and Z.~Wang, ``{ReAFM: A Reconfigurable Nonlinear Activation Function Module for Neural Networks},'' {\em IEEE Transactions on Circuits and Systems II: Express Briefs}, vol.~70, pp.~2660--2664, July 2023.

\bibitem{QForce-RL}
A.~Jha, T.~Dewangan, M.~Lokhande, and S.~K. Vishvakarma, ``{QForce-RL: Quantized FPGA-Optimized Reinforcement Learning Compute Engine},'' {\em 29th International Symposium on VLSI Design and Test}, July 2025.

\bibitem{Flex-SFU}
R.~Andri, E.~Reggiani, and L.~Cavigelli, ``{Flex-SFU: Activation Function Acceleration with Non-Uniform Piecewise Approximation},'' {\em IEEE Transactions on Computer-Aided Design of Integrated Circuits and Systems}, pp.~1--1, 2025.

\bibitem{TEAS}
Z.~Mei, H.~Dong, Y.~Wang, and H.~Pan, ``{TEA-S: A Tiny and Efficient Architecture for PLAC-Based Softmax in Transformers},'' {\em IEEE Transactions on Circuits and Systems II: Express Briefs}, vol.~70, pp.~3594--3598, Sept. 2023.

\bibitem{ShortcutFusion}
D.~T. Nguyen, H.~Je, T.~N. Nguyen, {\em et~al.}, ``{ShortcutFusion: From Tensorflow to FPGA-Based Accelerator With a Reuse-Aware Memory Allocation for Shortcut Data},'' {\em IEEE Transactions on Circuits and Systems I: Regular Papers}, vol.~69, pp.~2477--2489, June 2022.

\end{thebibliography}

\end{document}